\documentclass[12pt,thmsa,arabtex]{article}
\usepackage{graphicx}

\input{tcilatex}

\begin{document}

\begin{center}
\textbf{{\large A treatment of the quantum partial entropies in the
atom-field interaction with a class of Schr\"{o}dinger cat states}}

~

A.-S. F. Obada$^{1}$, H. A. Hessian$^{2}$ and Mahmoud Abdel-Aty$^{3,}$%
\footnote[4]{Corresponding author: abdelatyquantum@yahoo.co.uk}

$^1${\small Mathematics Department, Faculty of Science Al-Azhar University,
Nasr City, Cairo, Egypt}

$^2${\small Institut f\"{u}r Theoretische Physik, Universit\"{a}t Innsbruck,
Technikerstrasse 25, A-6020 Innsbruck, Austria }

$^{3}${\small Mathematics Department, Faculty of Science, South Valley
University, 82524 Sohag, Egypt }
\end{center}

\textbf{Abstract:} This communication is an enquiry into the circumstances
under which entropy and subentropy methods can give an answer to the
question of quantum entanglement in the composite state. Using a general
quantum dynamical system we obtain the analytical solution when the atom
initially starts from its excited state and the field in different initial
states. Different features of the entanglement are investigated when the
field is initially assumed to be in a coherent state, an even coherent state
(Schrodinger cate state) and a statistical mixture of coherent states. Our
results show that the setting of the initial state and the Stark shift play
important role in the evolution of the sub-entropies and entanglement.

instabilities, optical chaos and complexity and optical spatio-temporal
dynamics\}, {03.65.Ud} \{Entanglement and quantum nonlocality (e.g. EPR
paradox,Bell ?s inequalities, GHZ states, etc.)\}, {03.67.Hk} \{Quantum
communication\}

~

\section{Introduction}

The quantum entropy and entanglement is a vital feature of quantum
information. It has important applications for quantum communication and
quantum computation, for example, quantum teleportation, massive parallelism
of quantum computation and quantum cryptographic schemes [1-3]. Therefore,
it is very essential and interesting how to measure the entanglement of
quantum states. Recently much attention has been focused on the entanglement
of the field and atom when the system starts from a pure state [4-16]. Also,
in the context of the initial mixed state some studies have been reported
[17-20]. In this context it was shown that calculating the partial entropies
of the field or the atom can be used as an operational measure of the
entanglement degree of the generated quantum state. One finds that the
higher the entropy, the greater the entanglement. Starting from an initial
atom-field product state one can find perfectly entangled states between
field and atom at certain later times even for initial coherent states with
large photon number [4-8]. However, the time evolution of the field (atomic)
entropy reflects the time evolution of the degree of entanglement only if
one deals with a pure state of the system with zero total entropy.

\bigskip

The quest for proper entanglement measures has received much attention in
recent years [1,9]. From the identification and study of properties of such
measures a gain of insight into the nature of entanglement is expected. In
turn, their computation for particular states provide us with an account of
the resources present in those states. Because one needs to understand the
best way to benchmark states for quantum information protocols, here we
examine the quantum partial entropies in the atom-field interaction for more
general entangled states. From a practical point of view, an implementation
of the quantum entropy will be used to measure the entanglement degree when
the atom is assumed to be in its excited state and the field initially is in
a coherent state, superposition state and a statistical mixture of two
coherent states. As far as we are aware in the previous investigations, that
have dealt with the present problem, the initial system density matrix is
taken to be a product of two states of the factored form. The atom is often
taken to be in the excited pure state or mixed state and the radiation field
is taken to being a pure state density matrix. So, the present task is a
nontrivial issue, since we look at the mixed state entanglement from other
direction taking into account the entropy of the field not equal to the
entropy of the atom, in this case. To overcome such a difficulty, we employ
the quantum von-Neumann entropy to measure the entropy of the atom while a
numerical method will be used to calculate the quantum entropy of the field.

The material of this paper is arranged as follows. In section 2, we find the
exact solution of the system and write the expressions for the final state
vector at any time $t>0$. We investigate the quantum field (atomic) entropy
and the atom-field entanglement in section 3. Finally, numerical results and
conclusions are provided in section 4.

\section{The model}

The system we will consider here consists of a two-level atom interacting
with a single-mode quantized field via k-quanta processes. The Hamiltonian
in the rotating wave approximation [8,20], can be written as $(\hbar =1)$:
\begin{equation}
{\hat{H}}={\hat{H}}_{A}{+\hat{H}}_{F}{+\hat{H}}_{in}{,}
\end{equation}
where
\begin{eqnarray}
{\hat{H}}_{F} &=&\omega {\hat{a}}^{\dagger }{\hat{a},}  \nonumber \\
{\hat{H}}_{A} &=&\frac{\omega _{\circ }}{2}{\hat{\sigma}}_{z},  \nonumber \\
{\hat{H}}_{in} &=&{\hat{a}}^{\dagger }{\hat{a}}(\beta _{1}|g\rangle \langle
g|+\beta _{2}|e\rangle \langle e|)+\lambda ({\hat{a}}^{\dagger k}{\hat{\sigma%
}}_{-}+{\hat{a}}^{k}{\hat{\sigma}}_{+}),\quad
\end{eqnarray}
where $\omega $ is the field frequency and $\omega _{\circ }$ is the
transition frequency between the excited and ground states of the atom. We
denote by ${\hat{a}}$ and ${\hat{a}}^{\dagger }$ the annihilation and the
creation operators of the cavity field respectively. $\beta _{1}$ and $\beta
_{2}$ are parameters describing the intensity-dependent Stark shifts of the
two levels that are due to the virtual transition to the intermediate relay
level, $\lambda $ is the effective coupling constant, $\hat{\sigma}_{z}$ is
the population inversion operator, and $\hat{\sigma}_{\pm }$ are the ''spin
flip'' operators, with the detuning parameter $\Delta =\omega _{\circ
}-k\omega $.

Let us consider, the atom starting in its excited state $|e\rangle $, i. e.,
\begin{equation}
\rho _{0}^{a}=|e\rangle \langle e|,  \label{eq:a2}
\end{equation}
and we are going to assume that the initial single mode electromagnetic
field inside the cavity is in a superposition state of the kind:
\begin{equation}
\rho _{0}^{f}=\frac{1}{A}\left( |\alpha \rangle \langle \alpha
|+r^{2}|-\alpha \rangle \langle -\alpha |+r|\alpha \rangle \langle -\alpha
|+r|-\alpha \rangle \langle \alpha |\right) ,  \label{eq:a3}
\end{equation}
where $A=\left( 1+r^{2}+2r\exp (-2\alpha ^{2})\right) ,$with $\alpha $ real.
The parameter $r$ takes the values $-1$, $0$ and $1$, which corresponds to
an odd coherent state, a coherent state and an even coherent state,
respectively. While for certain classes of states a superpositions of
coherent states, methods solely based on linear optical elements like beam
splitters and photodetections could be found\textrm{\ [21],} an
implementation covering other classes of entangled states remains a
challenge.

Also we want to see how different would the behavior of the system be if the
input state is a statistical mixture of states $|\alpha \rangle $ and $%
|-\alpha \rangle $, i.e.,
\begin{equation}
\rho _{0}^{f}=\frac{1}{2}\left( |\alpha \rangle \langle \alpha |+|-\alpha
\rangle \langle -\alpha |\right) .  \label{eq:a31}
\end{equation}%
It is to be noted that when we put $r=0$ in equation (4) we get the same
result as in Ref. 13. It is expedient to expand the atom-field state in
terms of the dressed states:%
\begin{eqnarray}
|\Psi _{+}^{(n)}\rangle  &=&\sin \theta _{n}|n,e\rangle +\cos \theta
_{n}|n+k,g\rangle ,  \nonumber \\
|\Psi _{-}^{(n)}\rangle  &=&\cos \theta _{n}|n,e\rangle -\sin \theta
_{n}|n+k,g\rangle ,  \label{8}
\end{eqnarray}%
which are the eigenstates of the interaction Hamiltonian, where
\begin{eqnarray}
\hat{H}|s,g\rangle  &=&E_{0}|s,g\rangle ,\quad 0\leq s<k  \nonumber \\
\hat{H}|\Psi _{\pm }^{(n)}\rangle  &=&E_{\pm }^{(n)}|\Psi _{\pm
}^{(n)}\rangle ,  \label{6}
\end{eqnarray}%
with the eigenvalues $E_{0}$ and $E_{\pm }^{(n)}$
\begin{eqnarray}
E_{\pm }^{(n)} &=&\omega \left( n+\frac{k}{2}\right) +\frac{\omega _{0}}{2}++%
\frac{1}{2}\left[ n\beta _{2}+\beta _{1}(n+k)\right] \pm \mu _{n},  \nonumber
\\
E_{0} &=&\left( s\beta _{1}-\frac{\Delta }{2}\right) ,
\end{eqnarray}%
where
\begin{eqnarray}
\mu _{n} &=&\sqrt{\nu _{n}^{2}+\tau _{n}^{2}},  \nonumber \\
\nu _{n} &=&\frac{\Delta }{2}+\frac{1}{2}(\beta _{2}n-\beta _{1}(n+k)),
\nonumber \\
\tau _{n} &=&\lambda \sqrt{\frac{(n+k)!}{n!}}.  \label{9}
\end{eqnarray}%
$\mu _{n}$ is a modified Rabi frequency. The angle $\theta _{n}$ is given by
\begin{equation}
\theta _{n}=\sin ^{-1}\left( \frac{\tau _{n}}{\sqrt{(\nu _{n}-\mu
_{n})^{2}+\tau _{n}^{2}}}\right) .  \label{10}
\end{equation}%
The unitary operator $\hat{U}_{t}$ can be written as
\begin{eqnarray}
\hat{U}_{t} &=&\sum\limits_{n=0}^{\infty }\left\{ \exp (-itE_{+}^{(n)})|\Psi
_{+}^{(n)}\rangle \langle \Psi _{+}^{(n)}|+\exp (-itE_{-}^{(n)})|\Psi
_{-}^{(n)}\rangle \langle \Psi _{-}^{(n)}|\right\}   \nonumber \\
&&+\sum\limits_{s=0}^{k-1}\exp (-itE_{0})|s,g\rangle \langle g,s|.
\end{eqnarray}%
Despite being straightforwardly solvable in this way, the JC-model is
well-known for the fact that the time-evolution of most expectation values
is usually expressible only in series form. Having obtained the explicit
forms of the unitary operator $\hat{U}_{t}$, for the system under
consideration then the eigenvalues and the eigenfunctions can be used to
discuss many features concerning the field or the atom.

Bearing these facts in mind we find that the evolution operator $\hat{U}_{t}$
takes the next from
\[
\rho _{t}=\left(
\begin{array}{c}
\rho _{1} \\
\rho _{3}%
\end{array}%
\begin{array}{c}
\rho _{2} \\
\rho _{4}%
\end{array}%
\right) ,
\]%
where $\left( \rho _{i}\right) _{nm}=\langle n|\rho _{i}|m\rangle ,$ $%
i=1,2,3,4.$ $\left( \rho _{1}\right) _{nm}=A_{n}(t)A_{m}^{\ast }(t),$ $%
\left( \rho _{2}\right) _{nm}=A_{n}(t)B_{m-k}^{\ast }(t),$ $\left( \rho
_{3}\right) _{nm}=B_{n-k}(t)A_{m}^{\ast }(t),$ and $\left( \rho _{4}\right)
_{nm}=B_{n-k}(t)B_{m-k}^{\ast }(t).$ The coefficients $A_{n}(t)$ and $%
B_{n}(t)$ are given by
\begin{eqnarray}
A_{n}(t) &=&q_{n}C_{n}\exp [-i\lambda t\delta _{+}(n)]\left( \cos \lambda
t\mu _{n}-i\eta _{n}\frac{\sin \lambda t\mu _{n}}{\mu _{n}}\right) ,
\nonumber \\
B_{n}(t) &=&-iq_{n}C_{n}\nu _{n}\exp [-i\lambda t\delta _{+}(n)]\frac{\sin
\lambda t\mu _{n}}{\mu _{n}},  \nonumber \\
R^{2} &=&\sqrt{\beta _{1}/\beta _{2}},\quad \eta _{n}=\frac{\delta }{2}%
+\delta _{-}(n),\quad \delta =\frac{\Delta }{\lambda },  \nonumber \\
\delta _{\pm }(n) &=&\left\{
\begin{array}{c}
\frac{1}{2R}[n\pm R^{2}(n+k)], \\
0%
\end{array}%
\right.
\begin{array}{c}
when\ R\neq 0 \\
when\ \beta _{i}=0,%
\end{array}
\nonumber \\
&&
\end{eqnarray}%
where
\[
C_{n}=[\frac{1}{\sqrt{A}}(1+r(-1)^{n})],
\]%
for the initial condition (3), while for the initial condition (4) is
\begin{equation}
C_{n}=[\frac{1}{\sqrt{2}}(\delta _{i}+(-1)^{n}\delta _{j})],
\end{equation}%
with $\delta _{i}$, $\delta _{j}$ satisfying the two following condition, $%
(a)$ $\delta _{i}=\delta _{j}=(\delta _{i})^{2}=(\delta _{j})^{2}=1$, and $%
(b)$ $\delta _{i}.\delta _{j}=0$,
\begin{equation}
|\alpha \rangle =\sum_{n=0}^{\infty }q_{n}|n\rangle =\sum_{n=0}^{\infty
}e^{-\alpha ^{2}/2}\frac{\alpha ^{n}}{\sqrt{n!}}|n\rangle .  \label{eq:a4}
\end{equation}%
With the final state obtained, any property related to the atom or the field
can be calculated. Employing the reduced density operator for the atom or
the field, we investigate the properties of the entropies ($S_{a},S_{f}$)
and hence entanglement.

\section{Entropy and subentropy}

There is growing interest in the roles of nonadditive measures in quantum
information theory. Inadequacy of the additive Shannon von-Neumann entropy
as a measure of the information content of a quantum system has been pointed
out [22]. Also, there is a theoretical observation [23] that the measure of
quantum entanglement may not be additive. Despite the fact that the basic
idea of quantum entanglement was acknowledged almost as soon as quantum
theory was discovered, it is only in the last few years, that consideration
has been given to finding mathematical methods to generally quantify
entanglement. In the case of a pure quantum state of two subsystems, a
number of widely accepted measures of entanglement are known. However, the
question of quantifying the degree of entanglement for general mixed states
is still under discussion. Let us now briefly repeat some of the key
underlying definitions. The entropy $S$ of a quantum-mechanical system
described by the density operator ${\hat{\rho}}$ is defined as follows:
\begin{equation}
S=-Tr\{{\hat{\rho}}\ln {\hat{\rho}}\},  \label{eq:a8}
\end{equation}
where we have set the Boltzmann constant $K$ equal to unity. If ${\hat{\rho}}
$ describes a pure state, then $S=0$, and if ${\hat{\rho}}$ describes a
mixed state, then $S\neq 0$. Entropies of the atomic and field sub-systems
are defined by the corresponding reduced density operators:
\begin{equation}
S_{a(f)}=-Tr_{a(f)}\{{\hat{\rho}}_{a(f)}ln\rho _{a(f)}\}.
\end{equation}
Taking the partial trace over the field, the reduced atomic matrix can be
written as
\begin{eqnarray}
\rho _{t}^{a} &=&Tr_{f}(\rho ) =\left(
\begin{array}{c}
\rho _{ee} \\
\rho _{ge}%
\end{array}
\begin{array}{c}
\rho _{eg} \\
\rho _{gg}%
\end{array}
\right)  \nonumber \\
&=&\left(
\begin{array}{c}
\sum\limits_{n=0}^{\infty }|A_{n}(t)|^{2} \\
\\
\sum\limits_{n=0}^{\infty }B_{n+k}(t)A_{n}^{*}(t)%
\end{array}
\begin{array}{c}
\sum\limits_{n=0}^{\infty }A_{n}(t)B_{n+k}^{*}(t) \\
\\
\sum\limits_{n=0}^{\infty }|B_{n+k}(t)|^{2}%
\end{array}
\right) .
\end{eqnarray}
Thus we rigorously obtain the quantum atomic entropy in the following form
\begin{equation}
S(\rho _{t}^{a})=-\lambda _{+}^{a}(t)\log \lambda _{+}^{a}(t)-\lambda
_{-}^{a}(t)\log \lambda _{-}^{a}(t),  \label{22}
\end{equation}
where $\lambda _{i}^{a}(t)$ is given by
\begin{equation}
\lambda _{\pm }^{a}(t)=\frac{1}{2}\left\{ 1\pm \sqrt{\left( 2\rho
_{ee}(t)-1\right) ^{2}+4|\rho _{eg}(t)|^{2}}\right\} .  \label{23}
\end{equation}
In this case, the probability of finding the atom in its excited or ground
states are expressed as the diagonal element of the reduced atomic density
matrix, i.e.,
\begin{equation}
\rho _{ii}(t)=\langle i|\rho _{t}^{a}|i\rangle ,\qquad i=e,g
\end{equation}
and the off-diagonal element $\rho _{eg}(t)$ is given by
\begin{equation}
\rho _{ij}(t)=\langle i|\rho _{t}^{a}|j\rangle ,\qquad i=e,g.
\end{equation}
Taking the partial trace over the atomic system, we obtain the reduced
density operator in the form
\begin{equation}
\rho _{t}^{f}=tr_{A}\rho (t),  \label{16}
\end{equation}
with its ($\rho _{t}^{f})_{nm}$ element given by

\begin{equation}
(\rho _{t}^{f})_{nm}=A_{n}(t)A_{m}^{*}(t)+B_{n-k}(t)B_{m-k}^{*}(t).
\end{equation}
From this equation, it is difficult to obtain the eigenvalues of the reduced
density operator for the field, in this paper we will evaluate them
numerically.

\section{Results and conclusion}

We study the temporal behavior of the atom-field system in the JC-model for
the cavity-field prepared initially in different forms. As an example we may
consider a simple initial condition for the atom to be in the excited state
and the field in a coherent state or a superposition of the coherent state
or in a statistical mixture of two coherent states (equations (3) and (4)).
Among the family of mixed quantum mechanical states, special status should
be accorded to those for a given value of the entropy and have the largest
possible degree of entanglement. The reason for this is that such states can
be regarded as mixed-state generalizations of Bell states, the latter being
known to be the maximally entangled two-qubit pure states. Hence, this kind
of mixed states could be expected to provide useful resources for quantum
information processing. At this end, we have the plot of the quantum partial
entropies $(S_{a},S_{f})$ relative to these different initial states of the
atom-field, as a function of the scaled time $\lambda t/\pi ,$ taking into
account the two-photon process $(k=2),$ in order to investigate the Stark
shift effects.

\begin{figure}[tbph]
\begin{center}
\includegraphics[width=10cm]{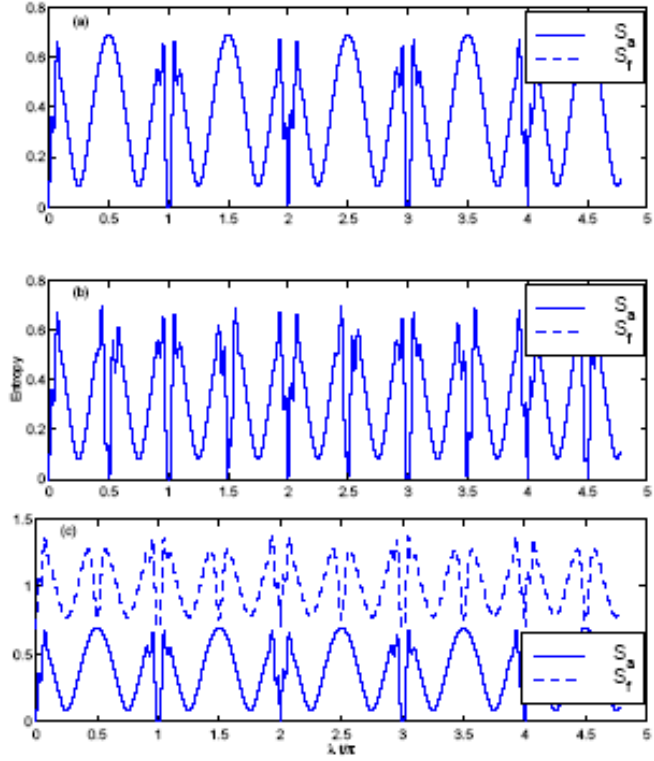}
\end{center}
\caption{The entropy for the atom $S_{a}(t)$ (solid line) and the field
entropy $S_{f}(t)$ (dashed line) as a function of the scaled time $\protect%
\lambda t/ \protect\pi$ of the particle initially prepared the excited state
and the field initially prepared in: $(a)$ a superposition state SS ($r=0$
(coherent state)), $(b)$ a superposition state ($r=1$ (even coherent state))
and $(c)$ a statistical mixture (MS) of coherent states $\mid \protect\alpha %
\rangle $ and $\mid -\protect\alpha \rangle $ $(\bar{n}=16)$.}
\end{figure}
We assume a fixed value of the initial mean number of quanta $\bar{n}=16$
and different values of Stark shift parameter $R$ (namely, $\beta _{1}=\beta
_{2}=0,$ i.e. in the absence of Stark shift in figure 1, $R=0.5$ in figure 2
and $R=0.3$ for figure 3). Furthermore, the detuning parameter is taken to
be zero, and in figure 1$a$ we set $r=0$ (coherent state), figure 1$b,$ we
set $r=1$ (even coherent state) and figure 1$c$ (a statistical mixture
state). In the absence of the Stark shift, it is observed that the quantum
field entropy and the quantum atomic entropy have the same values due to the
initial coherent state $r=0,$ (see figure 1a). This behavior is similar to
that obtained in the standard two-photon two-level systems obtained
previously (see for example [4-5]). It is observed that the entropy evolves
with a period $\pi /\lambda $, when $t=n\pi /\lambda ,n=0,1,2,...)$, the
quantum field entropy evolves to zero and the field is completely
disentangled from the atom, while for $t=(n+\frac{1}{2})\pi /\lambda $, it
evolves to the maximum value, and the field is strongly entangled with the
atom.

The situation is completely changed when we consider an even coherent state
i.e $r=1$. Although the quantum field entropy is still equal the atomic
entropy, the quantum entropy in this case has minimum values at $\lambda t=%
\frac{\pi n}{2}$, $(n=0,1,2,...)$ i.e at half of the revival time, also
instead of the two pre-minimum values observed in figure 1a we have here
only one pre-minimum value between each two consecutive minima. This effect
is due to the interference between the two coherent states in the
superposition, and can be understood by looking at the photon number
distribution of the initial field. Hence this signal gives a clear measure
of the remaining degree of coherence between the two components of the
Schrodinger cat state, while the signals present in both cases are due to
intrinsic revivals of each component individually. Because of this
enhancement it is possible to have the generation of well-defined Schr\"{o}%
dinger cat-like states during the evolution of the field in the two-photon
process case model [24]. We would like to remark that the approach to a pure
state at half of the revival time occurs in the ordinary JC-model [25], if
we start with the field in a pure state.
\begin{figure}[tbph]
\begin{center}
\includegraphics[width=10cm]{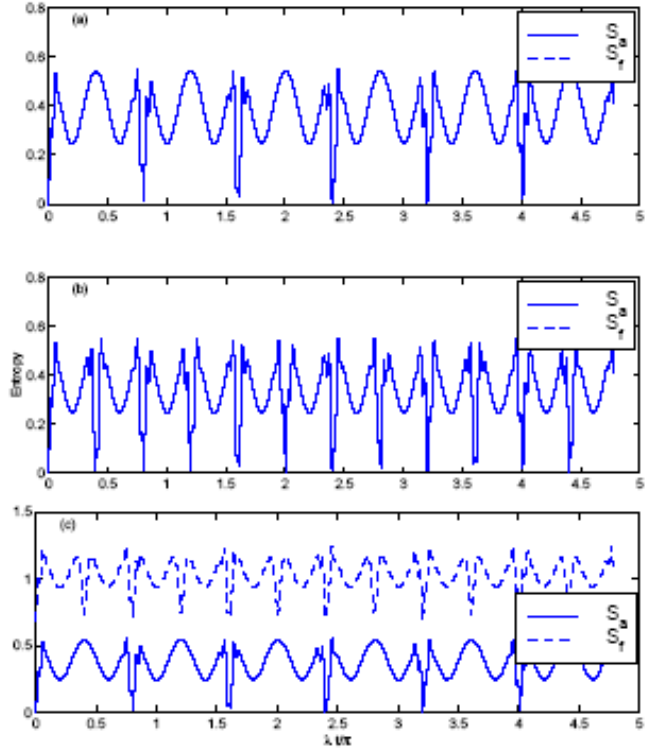}
\end{center}
\caption{The same as in figure 1 but with the Stark shift parameter $R=0.5$ }
\end{figure}
Let us now come to specific numerical examples to investigate the influence
of the statistical mixture on the evolution of the quantum field entropy and
the quantum atomic entropy (see figure 1c). An understanding of interaction
between an atom and an electromagnetic field has been possible in recent
years through the introduction of the statistical mixture state picture%
\textrm{\ [26].} By looking to the statistical mixture state one has a clear
physical understanding of what are the parameters involved in such
expression and what is going to neglect in order to go from a pure state to
a mixed state. On the other words, the state of the initial field is an
equally-weighted statistical mixture of two coherent states, which is a
special class of the Schrodinger cat state. In this case, as it has been
already discussed [20], the quantum field entropy $S_{f}(t)$ is greater than
the quantum atomic entropy $S_{a}(t)$ (see figure 1c), $S_{a}(t)$ reach its
maximum values at half of the revival time, while $S_{f}(t)$ evolves to
minimum values. We note a little deviation from ordinary Rabi oscillations,
due to the statistical mixture case. It is seen that entanglement evolves
depending on the initial preparation of atoms, however from figures 1a and
1c we see that the quantum atomic entropy has similar behavior in both
superposition and statistical mixture of coherent state.
\begin{figure}[tbph]
\begin{center}
\includegraphics[width=10cm]{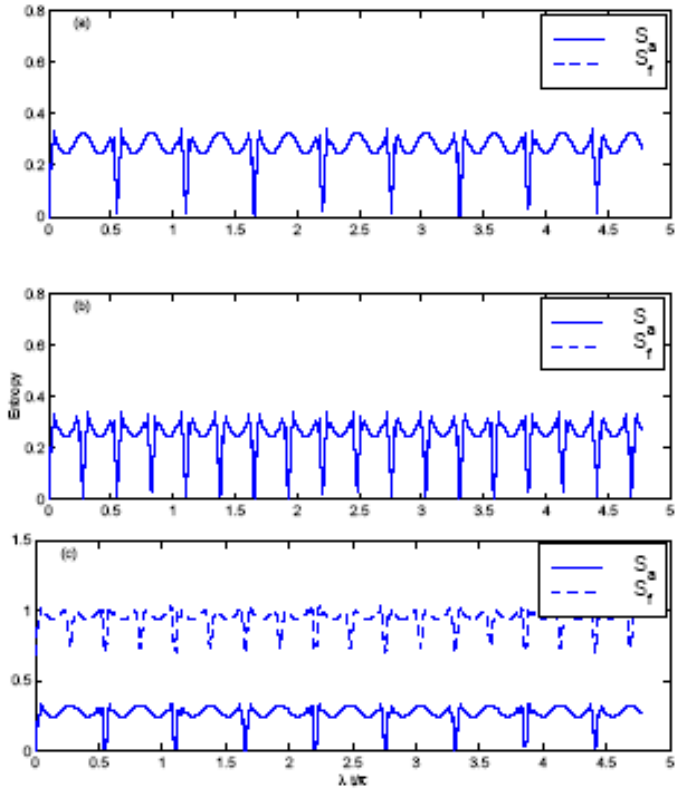}
\end{center}
\caption{The same as in figure 1 but with the Stark shift parameter $R=0.3$ }
\end{figure}

For comparison purposes, we have chosen to set some different values of the
Stark shift parameter R and the other parameters are the same as in figure
1. The outcome is presented in figure 2 (where $R=0.5$) and figure 3 (where $%
R=0.3$). We note a stronger modulation in the oscillations and a clear
departure from ordinary Rabi oscillations being verified as the Stark shift
parameter takes values far from the unity (see figures 2 and 3). Further
studies about such a situation have been carried out and are not displayed
here. It is interesting to refer here to the fact that the Stark shift
creates an effective intensity dependent detuning $\Delta _{N}=\beta
_{2}-\beta _{1}$ {[27]}. When $\beta _{2}=\beta _{1},\quad \Delta _{N}=0$,
in this case, the Stark shift does not affect the time evolution of the
quantum entropy. As is visible from the figures, the effects of the dynamic
Stark shift are more pronounced when $R$ deviates from unity. Interestingly,
when $R$ is decreases, the values of the maximum entropy are decreased.
Periodic models therefore may be more robust in this sense. Also, with
decreasing the parameter $R,$ the evolution period of the entropy as well as
the subentropies is decreases (see figure 3). The sensitivity becomes even
more clearly visible when we take small velues of the Strak shift parameter.
It is worth mentioning that the Strak shift effect has the same impact for
both the field entropy and the atomic entropy.

\bigskip

When the atomic entropy is calculated, we note that the terms involved are
of the forms $\left\langle U(t)\alpha |U(t)\alpha \right\rangle $ , $%
\left\langle U(t)\alpha |-U(t)\alpha \right\rangle $ and
$|\left\langle U(t)\alpha |U(t)\alpha \right\rangle |^{2}$ for the
case of the mixture. These terms in particular do not differs from
those of the case of the coherent state which has been discussed
earlier. Therefore the temporal
evolution of the entropy for the atomic system alone in the case of the Schr%
\"{o}dinger cat sate ($S$) mimics the evolution of the entropy in the pure
coherent state as can be seen from comparing figures (1c) and (1a). However
when we consider the entropy for the field we note that terms of the form $%
\left\langle \pm U(t)\alpha |\pm U(t)\alpha \right\rangle $ with all
combinations. These terms are the ones that appear in the case of
the superposition of the two states ($r=1$ in equation 4). Hence the
resemblance between the figures for the entropy of the field in the
mixed state of figure (1c) and the case of the initial superposed
states of figure (1b). This may demonstrate relevance of
investigating the entropies of the subsystems and their relation to
entanglement.

\bigskip

In summary, we have shown in this paper that the final analytical expression
of the composite density matrix along with its overlap matrix elements can
be used to obtain the quantum field entropy and the quantum atomic entropy.
This is accomplished by choosing to study the system in the representation
in which the marginal initial density matrices are assumed to be in a
coherent state, superposition states and statistical mixture states of two
coherent states. We present different numerical examples to elucidate the
effects of these different settings. Explicit computations are presented for
different values of the Stark shift parameter. Our results show that the
superposition of coherent states and Stark shift play an important role in
the evolution of the quantum entropies in the two-quanta JC-model. In the
coherent state, the quantum field (atom) entropy reaches its maximum values
at the half of the revival time and the field and the atom are strongly
entangled, while in the even coherent state, the entropies at the same time
evolve to the minimum values (zero) and the field and the atom are strongly
disentangled. In the statistical mixture state, ($S_{a}\neq S_{f}$), the
entropy for the atom reach to the maximum values at half of the revival
time, while $S_{f}$ evolves to its minimum values. The significant effect of
the Stark shift parameter appears when $R$ deviates from unity. The more $R$
deviates the more the two systems are weakly entangled.

\end{document}